# Fractal generalized zone plates

**Omel Mendoza-Yero[1,*], Mercedes Fernández-Alonso[1], Gladys Mínguez-Vega[1], Jesús Lancis[1], Vicent Climent[1], and Juan A. Monsoriu[2]**

[1] *GROC, Departament de Física, Universitat Jaume I, E12080 Castelló, Spain*
[2] *Departamento de Física Aplicada, Universidad Politécnica de Valencia, E-46022 Valencia, Spain*
*Corresponding author: omendoza@uji.es*

**Abstract:** The construction of fractal generalized zone plates (FraGZPs) from a set of periodic diffractive optical elements with circular symmetry is proposed. This allows us to increase the number of foci of a conventional fractal zone plate (FraZP), keeping the self-similarity property within the axial irradiance. The focusing properties of these fractal diffractive optical elements for points not only along but also in the close vicinity of the optical axis are investigated. In both cases analytical expressions for the irradiance are derived. Numerical simulations of the energetic efficiency of FraGZPs under plane wave illumination are carried out. In addition, some effects on the axial irradiance caused by the variation in area of their transparent rings are shown.

## 1. Introduction

The fractal structures play an important role in describing and understanding a large number of phenomena in several areas of science and technology [1]. Within the optical community, there is an increasing motivation to implement optical structures that exhibit fractal properties. Among them, it has been shown that optical fields derived from the diffraction of waves by fractal structures can originate self-similar on-axis irradiance profiles under monochromatic illumination [2, 3]. It means that, within a focal energy distribution some of its parts have the same shape as the whole irradiance profile. The above phenomenon appears when we dial with a fractal zone plate (FraZP) designed on the base of the well-known Fresnel zone plates (FZPs) [4, 5]. The focusing properties of the former optical elements were investigated in a theoretical manner, using the Fresnel approximation [2, 3], as well as from an experimental implementation with liquid-crystal displays [6, 7]. In the practice, they have been used to produce a sequence of focused optical vortices [8] or to achieve optical images with an extended depth of field and reduced chromatic aberration under white-light illumination [9].

In this manuscript, the construction of FraGZPs from a set of GZPs is proposed. By GZP we mean a circularly symmetric binary pupil, periodic in the squared radial coordinate, being the ratio between areas of the whole period and of its transparent part a positive integer number $\varepsilon$. It should be noted that the focusing properties of these GZPs under monochromatic and femtosecond illumination were extensively studied [10]. With the introduction of FraGZPs the versatility of FraZPs will be improved in a sense that the new elements can produce more foci. The position and peak height of several intense foci will be analytically determined. We find out that to increase the number of foci of a FraZP can find application, for instance, to trap and manipulate particles at different controlled levels by means of a spiral FraZP [8].

On the other hand, thanks to the diffractive nature of FraZPs, they can be used in optical regions where refractive optics is not available due to the strong absorption of materials, such as soft X-ray microscopy [11] or THz imaging [12]. To this end, it is essential to know the diffraction-limited resolution and/or the energetic efficiency of the fractal structures. Regarding this point, a novel study of the three dimensional light distribution of FraGZPs in the vicinity of the optical axis is included.

## 2. On-axis focusing properties of fractal generalized zone plates

The construction of a FraGZP from a set of GZPs is shown in Fig 1 for $\varepsilon = 4$. To determine the spatial distribution of their zones, the radii in the squared radial coordinate of transparent and opaque rings of each GZP are represented by consecutive open and filled line segments, forming a geometric bar. All bars are placed near each other and aligned to form a column of bar lines. Then, the amplitude transmittance in the squared radial coordinate of the binary pupil is obtained from the rotation the whole structure around one extreme. In this process, the open segments that coincide in position with filled ones are canceled. The remaining open segments transform, after the change of variable, into the transparent rings of the resultant binary pupil.

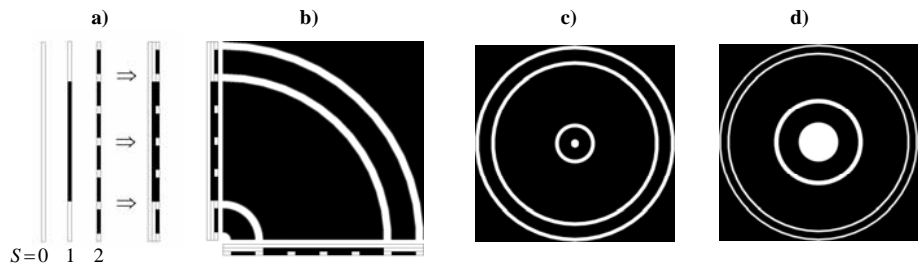

Fig. 1. Steps for construction of a FraGZP with $N = 2$ and $S = 2$ from GZPs with $\varepsilon = 4$, a) geometric bars, b) rotation process, c) pupil in $r^2$ and d) pupil in $r$.

It can be shown that the period of GZPs, normalized to $a^2$, is given by $p = \varepsilon /(\varepsilon N - (\varepsilon -1))^S$. After taking into account the dependence of $p$ on the parameter $\varepsilon$, the on-axis irradiance of a FraZP [2], can be generalized and rewritten in the form

$$I_{FraGZP}(u,N,S,\varepsilon) = I_{GZP}(u,N,S,\varepsilon) \prod_{i=1}^{S-1} \left[ \sin^2\left( N \frac{\varepsilon \pi u}{(\varepsilon N - (\varepsilon -1))^i} \right) \bigg/ \sin^2\left( \frac{\varepsilon \pi u}{(\varepsilon N - (\varepsilon -1))^i} \right) \right], \quad (1)$$

where $I_{GZP}(u,N,S,\varepsilon)$, denotes the on-axis irradiance originated by a GZP, with $N$ transparent rings and period $p$, that is illuminated with a monochromatic plane wave

$$I_{GZP}(u,N,S,\varepsilon) = 4\sin^2\left( \frac{\pi u}{(\varepsilon N - (\varepsilon -1))^S} \right) \sin^2\left( M \frac{\varepsilon \pi u}{(\varepsilon N - (\varepsilon -1))^S} \right) \bigg/ \sin^2\left( \frac{\varepsilon \pi u}{(\varepsilon N - (\varepsilon -1))^S} \right). \quad (2)$$

In Eq. (2) the number of transparent rings is expressed for convenience by the term $M$, so in this case $M = N$. The term $u = a^2 /(2\lambda z)$ is a normalized axial coordinate. The slowly oscillating right-hand term in Eq. (2) determines the energetic content of the different foci. In addition, the trigonometric quotient allows us to derive their axial locations. These are given by $z_n = pa^2 /(2\lambda n)$, with $n = 1, 2, ...$ and so on. An exception occurs when the equality $n = m\varepsilon$ is fulfilled, where $m$ is also a positive integer. In this case, the focus transforms into a phase singularity where the intensity vanishes [10]. The function given by Eq. (1) also achieves maximum values at the positions $z = z_n$. We refer to the foci associated with them as principal foci and to the remaining ones as secondary foci. The peak heights of the principal foci are assessed after solving the limit of $I_{FraGZP}(u,N,S,\varepsilon)$ as $z$ approaches $z_n$, whose value is $4N^{2S} \sin^2(n\pi/\varepsilon)$. This information is valuable for practical applications with fractal structures [6-8].

In order to analyze the on-axis focusing characteristics under monochromatic illumination of FraGZPs, six irradiance curves are plotted in Fig. 2 (middle-top). To do that, the parameters $N = 2$, $S = 2, 3$, $a = 10^{-2} m$, $\lambda = 780 nm$ and $\varepsilon = 2, 3$ and $4$ were substituted into Eq. (1). The self-similarity property is observed from two characteristic irradiance profiles with different fractal levels. That is, the patterns with $S = 3$ (top part of Fig. 2) are modulated versions of the corresponding with previous stage $S = 2$ plotted just at the middle.

The irradiance profile caused by the associated GZPs is shown in the bottom of Fig. 2 (only for $S = 2$). Theses curves were obtained from Eq. (2) after setting $M = (\varepsilon N - (\varepsilon -1))^S /\varepsilon + (\varepsilon -1)/\varepsilon$. The peak height of their principal foci are determined from Eq. (2) to yield $4M^2 \sin^2(n\pi/\varepsilon)$.

From the insets at the middle and bottom curves of Fig. 2 one can realize that a FraGZP is no more than a GZP with some missing transparent rings. Furthermore, a FraGZP with $S = 2$ may be considered as a particular case of a lacunar FraZP [3]. However, this does not apply to the case $S = 3$ or higher fractal orders.

After a visual inspection of Fig. 2 it is clear that the number of foci increases with increasing the parameter $\varepsilon$. Within a characteristic irradiance profile of a FraGZP, this number is approximately given by $(\varepsilon -1)(\varepsilon +1)^{S-1}$, for even value of the parameter $\varepsilon$. When $\varepsilon$ takes an odd value, it yields $(\varepsilon -1)\varepsilon^{S-1}$. The above expressions do not take into account irradiance peaks with relative low heights. Note that, the peak heights of some secundary foci, numerically determined from Eq. (1), cannot be disregarded because they have similar order of magnitude of that of the principal foci. From the comparison of the characteristic irradiance profiles of a FraGZP and its corresponding GZP one identifies the common foci. This link is suggested by Eq. (1) and Eq. (2).

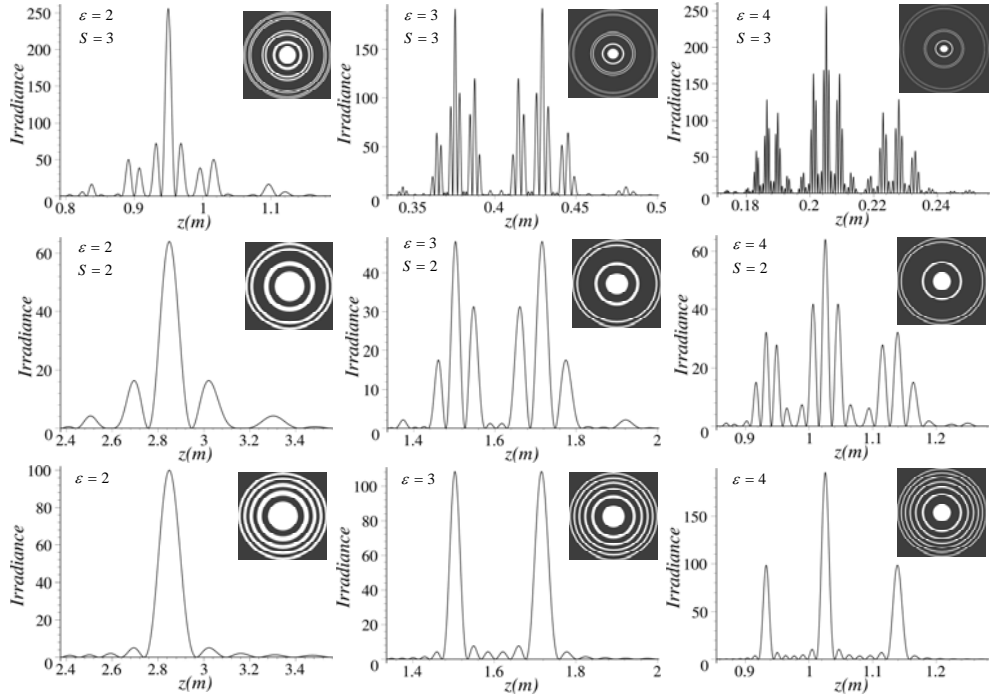

Fig. 2. Characteristic irradiance profiles of FraGZPs (top and middle) and GZPs (bottom).

## 3. Three-dimensional light distribution around the optical axis

The off-axis behavior of the diffracted field cannot be studied with the above formulation. To this end, an approximate analytical expression for the irradiance in terms of the first Rayleigh–Sommerfeld diffraction integral can be derived [13]. For points in the close vecinity of the optical axis, it yields

$$I(u,R) = \left\{\sum_{m=1}^{M}\left[J_0\left(4\pi\frac{ur_{im}R}{a^2}\right)\cos\left(2\pi\frac{ur_{im}^2}{a^2}\right) - J_0\left(4\pi\frac{ur_{om}R}{a^2}\right)\cos\left(2\pi\frac{ur_{om}^2}{a^2}\right)\right]\right\}^2 + \\ + \left\{\sum_{m=1}^{M}\left[J_0\left(4\pi\frac{ur_{im}R}{a^2}\right)\sin\left(2\pi\frac{ur_{im}^2}{a^2}\right) - J_0\left(4\pi\frac{ur_{om}R}{a^2}\right)\sin\left(2\pi\frac{ur_{om}^2}{a^2}\right)\right]\right\}^2 \quad (3)$$

In Eq. (3), $r_{im}$ and $r_{om}$ denote the inner and outer radii of the transparent rings which are given by $r_{im} = a[p(m-1)]^{1/2}$ and $r_{om} = a[p(m-(1-1/\varepsilon))]^{1/2}$, respectively, for a GZP. The radial coordinate in the output transversal plane is given by the variable $R$. The function $J_0(x)$ is the Bessel function of first kind with order zero and argument $x$. In general, Eq. (3) can be applied to simulate the off-axis irradiance originated by the diffraction of a plane wave through any binary amplitude pupils with circular symmetry. In the case of a FraGZP, attention must be paid to consider only the suffixes $m$ corresponding to the remaining clear zones resulting from the construction process, see Fig. 1. When $R = 0$, we can use Eq. (3) to plot the curves of Fig. 2.

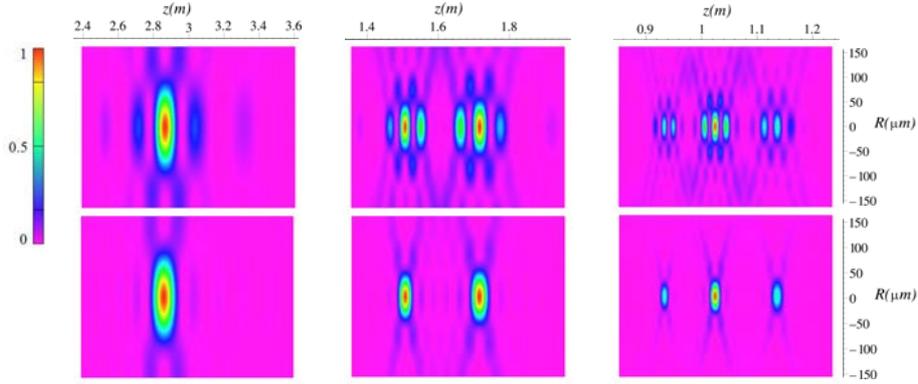

Fig. 3. Normalized off-axis irradiance profiles of FraGZPs (top) and GZPs (bottom) given in the middle-bottom parts of Fig. 2.

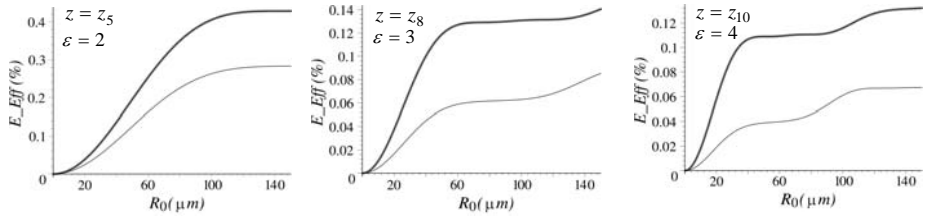

Fig. 4. Energetic efficiency of FraGZPs (thin curves) and GZPs (thick curves) at the planes $z = z_5$ (for $\varepsilon = 2$), $z = z_8$ (for $\varepsilon = 3$), and $z = z_{10}$ (for $\varepsilon = 4$). In all cases $S = 2$.

To be consistent with the previous analysis, we focus our attention on the off-axis behavior of the irradiance caused by pupils depicted in Fig. 2, specifically those characterized by $S = 2$ together with GZPs. In Fig. 3 the complex structure of foci near the optical axis is shown. From the three-dimensional light distribution, one can compare the ability of FraGZPs and corresponding GZPs to concentrate the energy around the optical axis. This type of plotting is important for microscopy, micromachining or lithography where there has been growing interest in the use of zone plates [14, 15].

If we look carefully at light distributions in Fig. 3, it can be noticed that transversal width of the principal foci are almost the same for a fractal structure and its corresponding GZP, but the axial width is reduced with the former pupil. The later effect is maybe more clearly seen on Fig. 2. This behavior suggests that a FraGZP could be used as a lens to enhance the resolving power of a GZP, keeping in mind that when using a fractal pupil the peak height is lower. To give a numerical example, we assessed the full width at half maximum in the axial direction of the foci at $z = z_7$ for the pupils in Fig. 2 with $\varepsilon = 3$ and $S = 2$. In this case, the ratio between the corresponding values yields $0.87$.

Within this context, the energetic efficiency of the binary pupils at any transversal plane $z$ is determined. We evaluate the expression $E\_eff(R_0) = 2\pi \int_0^{R_0} I(u, R) R dR / (\pi a^2)$ that gives us the ratio between the energy transmitted by a pinhole of radius $R_0$ at the output plane $z$ and the energy of the incident plane wave within the pupil area $A = \pi a^2$. The results for the binary pupils used in Fig. 3 at three principal planes $z = z_n$ are shown in Fig. 4, where the radius $R_0$ varies from $0$ to $150 \mu m$. Because of the absence of some rings, FraGZPs exhibit less efficient than corresponding GZPs. This difference is reduced as $\varepsilon$ tends to $2$, see insets in Fig. 2. The calculus of the energetic efficiency of binary pupils is a necessary requirement for many experiments, as the generation of non-linear effects, where FZPs have been successfully applied [16, 17]. Here, it should be pointed out that Fractal kinoform lenses are currently under development to improve the diffraction efficiency of fractal structures [18].

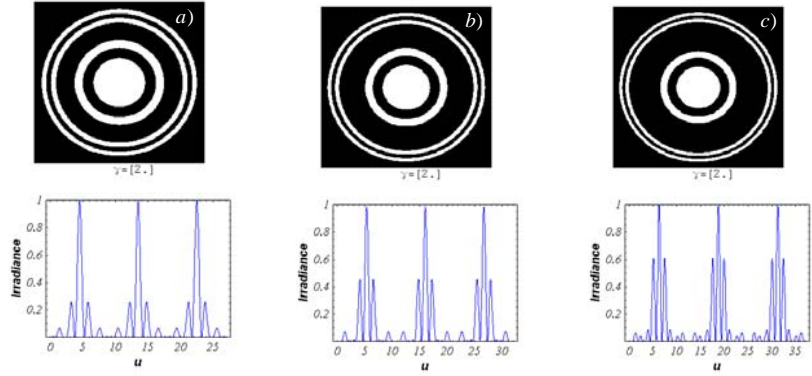

Fig. 5. Movies of the evolution with $\gamma$ of the normalized on-axis irradiance of FraGZPs as a function of the variable $u$ with a) $\varepsilon = 2$ (722 kB), b) $\varepsilon = 3$ (742 kB), and c) $\varepsilon = 4$ (750 kB). In all cases $N = 2$ and $S = 2$.

Finally, we study how the on-axis irradiance profile develops when the area of the transparent rings is modified. This is useful to get a deep insight into the focusing features of FraGZPs. Here, the period of the binary pupil is fixed, and consequently the inner radii $r_{im} = a[p(m-1)]^{1/2}$ too. Then, the outer radii are displaced all together in accordance with the expression $r_{om} = a[p(m-(1-1/\gamma))]^{1/2}$, where $\gamma \geq 1$ is a real number. Note that the pupil radius is no longer a constant but it changes with $\gamma$ as $r_{oM}^2 = a^2[1 - p(1-\varepsilon/\gamma)/\varepsilon]$. In this conditions, we use the Eq. (3) for $R = 0$ to make three animation movies of the irradiance evolution from $\gamma = 1$ to $\gamma = 6$ when $\varepsilon = 2$ (Fig. 5a), $\varepsilon = 3$ (Fig. 5b), and $\varepsilon = 4$ (Fig. 5c). At the same time, the transformation of the binary pupils as $\gamma$ changes is also shown. After looking at the animations, it is possible to see characteristic irradiance shapes, when $\gamma$ takes positive integer values. In these cases, the form of a characteristic profile depends on the particular value of $\gamma$, whereas the relative peak height of their foci on the period $p$.

## 4. Conclusions

FraGZPs constructed from periodic diffractive optical elements were introduced. The analytical expressions for the on-axis irradiance as well as for the position and height of the principal foci were obtained. The on-axis irradiance curves are related to the ones of corresponding GZPs by common foci, see Fig. 2. An approximate analytical expression for the irradiance in the close vicinity of the optical axis was derived. This allows analyzing the off-axis structure of foci determining, for instance, their transversal width and the energetic efficiency of FraGZPs. Note that, in the particular case of $\varepsilon = 2$, the on-axis irradiance function given by Eq. (1) reduces to that of a conventional FraZP. We also realize that a similar study for FraGZPs could be carried out with a different approach [19].

## Acknowledgements

This research was funded by the Conselleria de Empresa, Universitat i Ciència, Generalitat Valenciana, Spain, under the projects GV/2007/128 and GV/2007/239. We acknowledge partial support from the Consolider Programme SAUUL CSD2007-00013 and the Ministerio de Ciencia e Innovación (grant DPI2008-02953), Spain. Omel Mendoza-Yero thanks the "Convenio UJI-Fundació Caixa Castelló (Bancaixa)" for covering costs of the research. Juan A. Monsoriu also acknowledge the financial support from the Universidad Politécnica de Valencia (PAID-05-07).